\documentclass[12pt]{iopart}

\usepackage{iopams}

\expandafter\let\csname equation*\endcsname\relax

\expandafter\let\csname endequation*\endcsname\relax

\usepackage{amsmath}

\usepackage{graphicx}
\usepackage{float}
\usepackage{cite}
\usepackage{hyperref}
\usepackage{subcaption} 
\usepackage{centernot}
\makeatletter
\newcommand{\mainmatter}{%
  \setcounter{footnote}{0}%
  \patchcmd{\@makefntext}{\fnsymbol}{\arabic}{}{}%
  \patchcmd{\@thefnmark}{\fnsymbol}{\arabic}{}{}%
  \def\@makefnmark{\textsuperscript{\arabic{footnote}}}%
}
\makeatother
\begin{document}

\title[]{Pedagogical study of the image of a magnetic dipole in front of a superconducting sphere}

\author{%
Hemansh Alkesh Shah$^{1}$, Kolahal Bhattacharya$^{2}$
\footnote{Author to whom any correspondence should be addressed: kolahalbhattacharya@sxccal.edu}
}

\address{$^{1}$Indian Institute of Science, Bangalore, KA-560012 (India)\\
$^{2}$St. Xavier's College (autonomous), Kolkata, WB-700016 (India)}
\ead{$^{1}$hemansha@iisc.ac.in, $^{2}$kolahalbhattacharya@sxccal.edu}

\vspace{10pt}
\begin{indented}
\item[]
\end{indented}

\begin{abstract}
    The method of images to solve certain electrostatic boundary-value problems is taught worldwide in undergraduate-level physics courses. Though it is also possible to employ this technique for solving magnetostatic boundary value problems, examples of this usage are rarely found in textbooks or physics pedagogy literature. In particular, the problem of finding the field due to a magnetic dipole kept in front of a superconducting sphere is an interesting one, because (i) it helps the students to compare with the grounded conducting sphere image problem in electrostatics, and (ii) offers a greater degree of difficulty since the source is a dipole (vector), rather than an electric charge (scalar). The present work demonstrates an intuitive way of solving the problem. The case in which the source dipole is oriented radially with respect to the sphere is solved with a single dipole image. In the case of the transverse orientation of the source dipole, we model the dipole as a current loop. Then, we find the image of the radial and transverse current elements that satisfy the boundary conditions. Then, we show that this method can be used to deduce the form of the image dipoles when the dipole is oriented in the transverse direction. This method is very much intuitive and accessible for undergraduate-level students. 
\end{abstract}

%
\section{Introduction}\label{Sec1}
The method of images is an elegant technique to solve electrostatic boundary value problems. One can calculate the field or potential when a known charge distribution is placed near a surface (for example, a conducting body). The boundary condition on the surface is given, but the charge induced on the surface is unknown. The tricky job is accomplished by asserting that the field can be expressed as a sum of the field due to the known charge distribution, as well as that due to a simplified image charge distribution which effectively plays the role of the induced charge. We do not need to worry about the surface any more. The uniqueness theorem guarantees that the potential found by this technique is unique. The method was introduced by Lord Kelvin in 1848 in one paper published in Cambridge and Dublin Mathematical Journal~\cite{thomson1846cambridge}. It is taught nowadays in the undergraduate level course on electrostatics~\cite{griffiths2005introduction, greiner2012classical}. Apart from that, many interesting pedagogic applications~\cite{santos2004electrostatic, bhattacharya2011analogy} and research-level applications have been discussed in the  literature~\cite{norris1995charge, lin2006theoretical, redzic2011extension}. 

In the realm of magnetostatics, the method of images has been used to explain magnetic levitation near the surface of a superconductor~\cite{arkadiev1947floating}. Smythe~\cite{smythe1950static} (1950) used the concept of image currents to solve boundary value problems in magnetostatics (section 7.23: Current Images in Plane Face). In recent times, a few pedagogical articles used the concept to throw light on magnetic levitation near planar superconducting surfaces~\cite{saslow1991superconductor, badia2006understanding}. However, one does not find pedagogical examples of the method in magnetostatics in common textbooks. This is perhaps due to the fact that compared to the conductors, superconductivity is a relatively recent development in physics. The recent book by Matsushita~\cite{matsushita2021electricity}, however, contains a few simple examples of the applications of the method of images involving superconductors.  

Here let us mention that a formal introduction to superconductivity can be found in the textbook by Tinkham~\cite{tinkham2004introduction}. For temperatures $T$ below the critical temperature $T_C$ ($T<T_C$), the magnetic field ${\bf B}$ vanishes inside the superconductor, i.e. $B=0$. This is called the Meissner effect~\cite{meissner1933neuer}. When an external magnetic field ${\bf H}$ is applied, a superconductor expels the magnetic field from the interior by setting up an electric current at the surface. The surface current creates a magnetization that cancels the external magnetic field inside the bulk of the superconductor in the case of type-I superconductors. The magnetic field does not abruptly reduce to zero at the surface of a real superconductor. It does so over a small length scale called the London penetration depth. For its calculation, the interested reader is referred to the book by Tinkham~\cite{tinkham2004introduction}. In this work, we assume an ideal, Type-I superconductor with a vanishingly small London penetration depth.

Coming back to the discussion of magnetic images, a pedagogically more interesting case would be to calculate the magnetic field at the test point, when a magnetic dipole (or an equivalent current distribution) is placed in front of a spherical surface as shown in the following figure. 
\begin{figure}[H]
    \centering
    \includegraphics[width=0.4\linewidth]{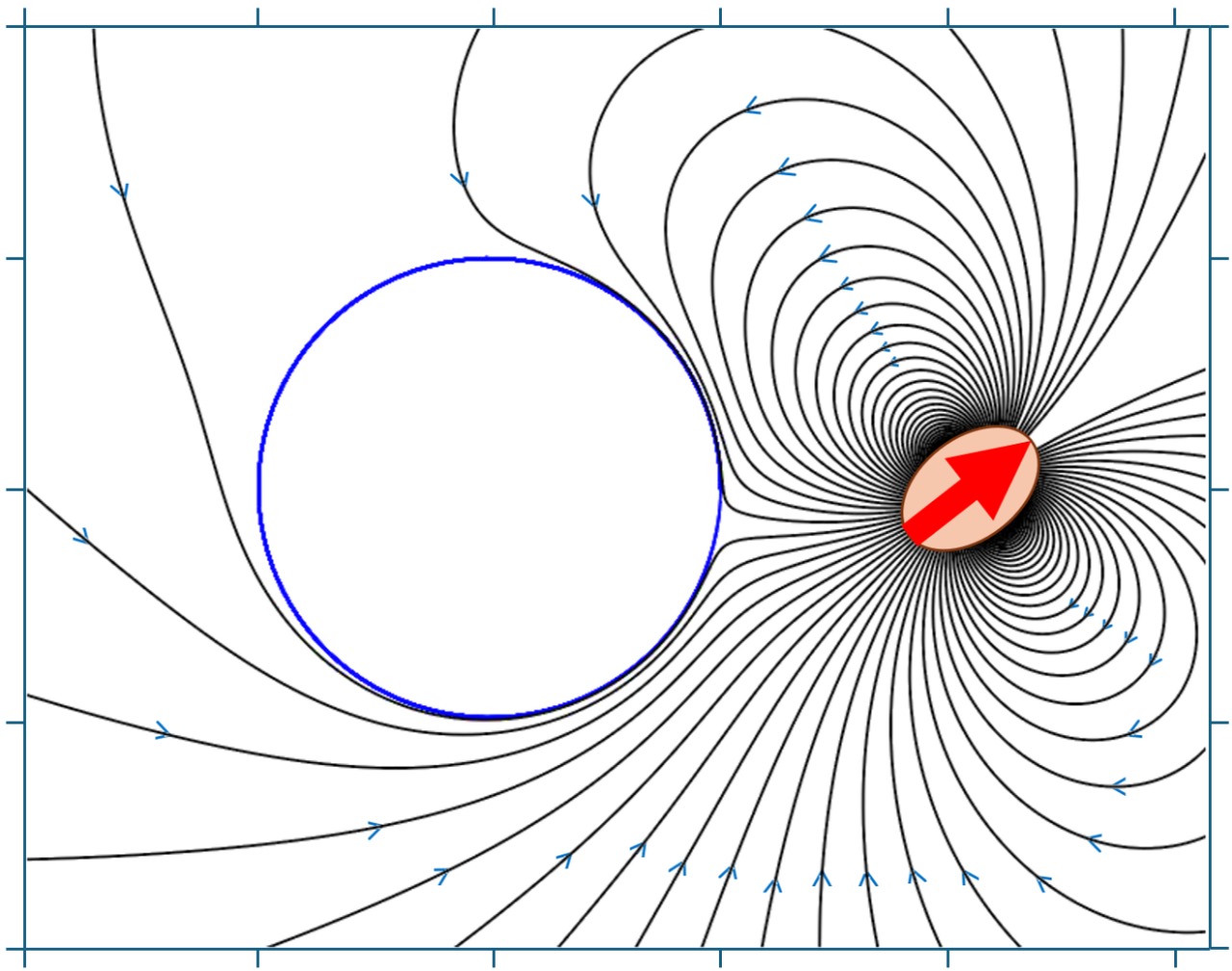}
    \caption{Field lines of a magnetic dipole (red colour) of arbitrary orientation placed in front of a superconducting sphere (depicted in blue). The field lines are parallel to the interface just outside the sphere. The magnetic field inside the sphere is zero.}
\end{figure}
Since a dipole of arbitrary orientation can be resolved into radial and transverse components, one can approach the problem systematically. Lindell analysed the case, where a current loop is placed in front of a material sphere, such that its axis is radially oriented relative to the sphere~\cite{lindell1993image}. He used an image current loop to solve the problem. In the limit where the size of the current loop is small, one can talk about a point magnetic dipole placed in front of a sphere pointing in the radial direction. The same technique was used by Sezigner et al.~\cite{sezginer1990image} in the context of a superconducting sphere. Rezdic~\cite{redvzic2006image, redvzic2006extension} showed that the it is possible to replace the superconducting sphere with a permeable sphere with an embedded current loop which coincides with the image current for the superconducting sphere case. Yang~\cite{yang1998interaction} and Coffey~\cite{coffey2000levitation} dealt with the same problem to find levitating force, without any reference to magnetic image currents. They matched the boundary conditions of the magnetic scalar potential due to the source and the image dipoles to solve the problem. However, the case of transverse orientation of the magnetic dipole (or current loop) was not covered in these works.

A systematic way to approach this problem is shown by Lin~\cite{lin2006theoretical} who separately considers the cases of the radial and transverse orientations. He treats a magnetic dipole as two closely spaced magnetic monopoles of opposite signs. Then, he solves the image problem for a single magnetic monopole and finally uses the principle of superposition to arrive at the correct answers. Other notable works on dipoles with transverse orientation came from Coffey~\cite{coffey2002london, coffey2002levitation}. However, like the radial orientation case, these works are also concerned with the levitating force but not with the image dipole configuration. Hence, the only consolidated solution to the problem using the method of images is the work by Lin~\cite{lin2006theoretical}. 
In many ways, this problem is an instructive tool for physics students. First of all, this problem gives an option for a comparative study between the application of the image method in electrostatics and magnetostatics. It also offers a higher degree of complication since the source is represented by a vector, whereas the standard electrostatic problem involves only a scalar charge. In principle, one can also solve the problem of an electric dipole placed in front of a grounded conducting sphere by the method of images~\cite{santos2004electrostatic}. This exercise, therefore, could give a great case-by-case comparison between the application of the method in electrostatics and magnetostatics. There is another simpler example of the use of the method in magnetostatics. Let us consider a straight infinite wire carrying constant current, placed at a small distance from an infinite superconducting plane; or a magnetic dipole placed over an infinite superconducting plane~\cite{saslow1991superconductor}. But these problems perhaps do not involve many interesting features of the more generalised problem of a sphere.

Despite Lin~\cite{lin2006theoretical} giving a systematic development of the application of the method of images in typical textbook-type problems and its conceptual appeal, we think this work has little pedagogical value, for many reasons. First of all, Lin uses magnetic scalar potential, a quantity that physics students are usually discouraged to use, due to the subtlety of the topological nature of the region where they could be applicable. Earlier Coffey~\cite{coffey2002levitation} justified the usage of scalar potential ``by the absence of electric current outside of the superconducting sphere''. This statement does not explicitly state the nature of the current. In this problem the magnetic scalar potential can be used, because we have a bound (induced) current on the sphere and the free current is zero. This would imply $\nabla \times{\bf H}={\bf J}_{free}={\bf0}$. In addition, we must consistently keep ourselves in a simply-connected region. Lin considers the region outside the sphere and gives the expressions for magnetic scalar potential due to the source monopole [Eq.(2) in~\cite{lin2006theoretical}] and that due to the image monopole's distribution [Eq.(3) in~\cite{lin2006theoretical}]. After that, it is stated without justification that the scalar potential due to the image monopole can be expressed as the difference between a monopole term and a line-monopole density term [Eq.(7) in~\cite{lin2006theoretical}]. This step appears ad-hoc from a pedagogical viewpoint. Finally, we note that using
magnetic monopole is a matter of convenience. One cannot adopt this as a physical theory, because monopoles have never been located in the universe, through several experiments. For these reasons, it is perhaps worthwhile to look for alternative pedagogical approaches to solve the problem. If we want to look beyond the magnetic scalar potential, a safe starting point would be to start with the boundary conditions matching the magnetic field ${\bf B}$ or the vector potential ${\bf A}$. For example, the normal component of the magnetic field must be zero at the surface:
\begin{equation}\label{eq1}
    B^\perp=0    
\end{equation}
due to the Meissner effect. We can use this condition to deduce the image configuration when the source magnetic dipole is oriented radially. The boundary condition on ${\bf A}$, as pointed out in Chapter 7 of~\cite{matsushita2021electricity}, is that if the superconducting region is simply-connected, then the vector potential can be assumed to be constant in this region:
\begin{equation}\label{eq2}
{\bf A}={\bf C}
\end{equation}
-where ${\bf C}$ denotes a constant vector. The vector potential is continuous across the surface. We can use this condition when the source dipole is oriented transversely.
\\
\section{The image of a radial dipole on a superconducting sphere}\label{Sec2}
The following figure~\ref{fig1} shows the configuration when a point magnetic dipole ${\bf m_1}$ is placed near a superconducting sphere of radius $a$. The sphere is centred at the origin of the coordinates $O$ and the dipole is located at a distance $d_1(>a)$ from $O$, pointing radially outwards. Due to the Meissner effect~\cite{meissner1933neuer}, the magnetic field inside the sphere vanishes. The dipole may also be thought of as a tiny current loop with area $A\rightarrow0$, carrying anti-clockwise current (when viewed from the top) $I\rightarrow\infty$, such that $|{\bf m_1}|=IA$ is finite. This dipole induces a surface current of unknown functional form on the sphere. The total magnetic field outside the sphere is the field due to the dipole plus the field due to the unknown surface current. How to find the field and the surface current?
\begin{figure}[ht]
\centering
\includegraphics[width=7.5cm, height=8.0cm]{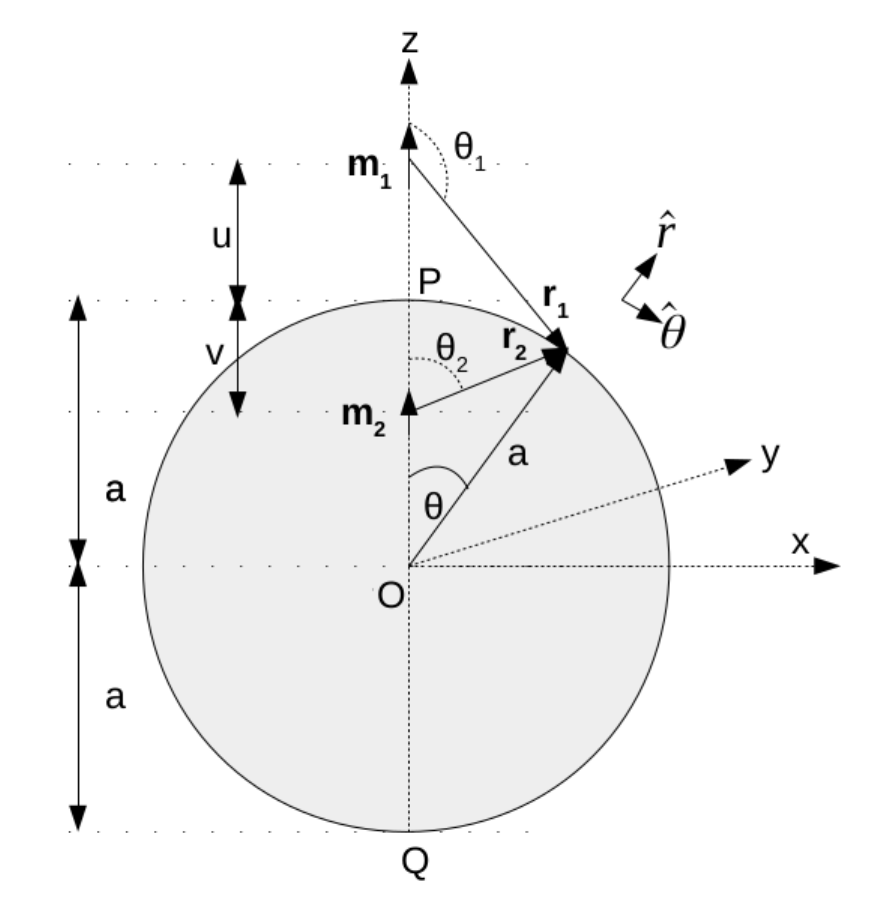}
\caption{Magnetic dipole placed radially in front of a superconducting sphere (shaded sphere). Magnetic field $B=0$ inside the sphere. In the figure, $d_1= a+u$.}
\label{fig1}
\end{figure}

The method of images asserts that this problem could be solved by assuming that there exists a distribution of the image dipole inside the sphere, which together with the real dipole, satisfies the boundary conditions at the location of the spherical surface. We shall assume that a point image dipole ${\bf m_2}$, pointing radially outwards, located at a distance $d_2$ from the origin $O$ is sufficient to address the problem. The relevant boundary condition is given by Eq.\eqref{eq1}. The existence of a unique real solution should follow if this is true.

In the scientific literature on the method of images in electrostatics, it has been illustrated previously that an analogy between the grounded conducting sphere image problem and optical image formation by the spherical mirror can be observed if the distances to the real and image dipoles are measured from the pole $P$, instead of from $O$~\cite{bhattacharya2011analogy}. We shall employ the same idea (object distance $u$ and image distance $v$) in this problem while implementing boundary conditions. Note that $d_1=u+a$ and $d_2=a-v$.

Let us investigate the boundary condition of a magnetic field at $P$, the pole, where the real and image dipoles will have only the radial components of the field. The contribution of the real dipole (${\bf m_1}$) placed outside the sphere to the total field is
\begin{equation}\label{eq3}
{\bf B}^{source}(P)=\frac{\mu_0{m_1}}{4\pi u^3}(2\cos\pi(-\hat {z})+\sin\pi(-\hat{x}))=\frac {2\mu_0{m_1}}{4\pi u^3}\hat{z}
\end{equation}
Similarly, the contribution to the field at $P$ from the image dipole ${\bf m_2}$ is
\begin{equation}\label{eq4}
{\bf B}^{image}(P)=\frac{\mu_0{m_2}} {4\pi v^3}(2\cos(0)(\hat{z})+\sin(0)(\hat{x}))=\frac{2\mu_0 {m_2}}{4\pi v^3}\hat{z}    
\end{equation}
The boundary condition at $P$ (given in Eq.\eqref{eq1}) translates to:
\begin{equation}\label{eq5}
 \frac{m_1}{u^3}+\frac{m_2}{v^3}=0
\end{equation}

Next, let us investigate the magnetic field boundary condition at a point $Q$ that is diametrically opposite to the pole $P$. There, the contribution to the field from the dipole ${\bf m_1 }$ is: 
\begin{equation}\label{eq6}
{\bf B}^{source}(Q)=\frac{\mu_0{m_1}}{4\pi(u+2a)^3}(2\cos\pi(-\hat{z})+\sin\pi(-\hat{x}))=\frac{2\mu_0{m_1}}{4\pi(u+2a)^3}{\hat{z}}    
\end{equation}
Similarly, the contribution to the field from the image dipole ${\bf m_2}$: 
\begin{equation}\label{eq7}
{\bf B}^{image}(Q)=\frac{\mu_0{m_2}}{4\pi(2a-v)^3}(2\cos\pi(-\hat{z})+\sin\pi(-\hat{x}))=\frac{2\mu_0{m_2}}{4\pi(2a-v)^3}(\hat{z})
\end{equation}
The last two equations lead to the following boundary condition at $Q$:
\begin{equation}\label{eq8}
 \frac{m_1}{(u+2a)^3}+\frac{m_2}{(2a-v)^3}=0
\end{equation}
From Eq.~\ref{eq5}, we express $m_1$ and $m_2$ in terms of $u$, $v$, $a$, and a non-zero constant, say, $k$:
\begin{align}\label{eq9}
    \frac{m_1}{u^3}&=-\frac{m_2}{v^3}=k\nonumber\\
\implies m_1&=u^3k, m_2=-v^3k
\end{align}
Using Eq.~\ref{eq8} and Eq.\eqref{eq9}, we deduce that:
\begin{align}\label{eq10}
    \frac{u^3k}{(u+2a)^3}&=\frac{v^3k}{(2a-v)^3}\nonumber\\
\implies\frac{u}{u+2a}&=\frac{v}{2a-v}\nonumber\\
\implies(2au-2av)&=2uv\nonumber\\
\implies \frac{1}{u}-\frac{1}{v}&=-\frac{1}{a}
\end{align}
Eq.~\eqref{eq10} is the mirror equation in this problem. Notice that Eq.(10) also means that $v=au/(a+u)=au/d_1$. This is the same as the corresponding mirror equation observed in~\cite{bhattacharya2011analogy} in the grounded conducting sphere image problem. The focal length $a$ denotes the distance from the pole $P$ where the image dipole would be located when the real dipole distance $u\rightarrow\infty$. The focal length equals the radius of the sphere, just as in electrostatics. The image distance and the focal length come with a negative sign because they are in a direction opposite to that of the real dipole distance following the trend used in geometrical optics. Only, in this case, the linear magnification is different. From Eq.~\ref{eq9} and Eq.\eqref{eq10}, we see:
\begin{equation}\label{eq11}
 m_2=-m_1\left(\frac{1}{u^3}\frac{a^3u^3}{(a+u)^3}\right)=-\frac{a^3}{d_1^3}{m_1}
\end{equation}
This also shows that the image dipole must be oriented in the radially inward direction. Though the solution was found without considering the possible dependence on $\theta$, we can check that the solution satisfies the boundary conditions even when $\theta\neq0$. For example, at an arbitrary point $(a,\theta)$ on the sphere, the radial component of the total magnetic field should vanish. To verify this, we must express the magnetic fields due to the dipoles in terms of $\theta$. First, we consider the dipole ${\bf m_1}$. Referring to figure~\ref{fig1}, the magnetic field due to the source dipole ${\bf m_1}$ and the image dipole are given by~\cite{griffiths2005introduction}:
\begin{align}\label{eq12}
    {\bf B}^{source}(a,\theta)=\frac{\mu_0 m_1}{4\pi r_1^3}(2\cos\theta_1{\hat{\bf r}_1}+\sin\theta _1 \hat{\bf\theta}_1)\nonumber\\
    {\bf B}^{image}(a,\theta)=\frac{\mu_0 m_2}{4\pi r_2^3}(2\cos\theta_2{\hat{\bf r}_2}+\sin\theta_2\hat{\bf\theta}_2)
\end{align}
where $(\theta_1,\theta_2)$ and $({\bf r_1}, {\bf r_2})$ are defined in figure~\ref{fig1}. We must express these in terms of the spherical polar coordinates $(a,\theta,\phi)$ centred at $O$. Note that $\hat{\bf r}_1\times\hat{\bf\theta}_1$ points into the page, in the same direction as $\hat{\bf r}\times\hat{\bf \theta}$. The radial component of the field with respect to the coordinate system at $O$ is given by:
\begin{equation}\label{eq13}
    {\bf B}^{source}(a,\theta)\cdot\hat{\bf r}=\frac{\mu_0m_1}{4\pi r_1^3}\left(2\cos\theta_1(\hat{\bf r}_1\cdot\hat{\bf r})+\sin\theta_1 (\hat{\theta_1}\cdot\hat{\bf r})\right)
\end{equation}
We need to express $\cos\theta_1$, $\sin\theta_1$, $\hat{\bf r}_1\cdot\hat{\bf r}$, as well as $\hat{\theta_1}\cdot\hat{\bf r}$ in terms of $\theta$ to make progress. We first find out $\hat {\bf r}_1\cdot\hat{\bf r}$. We start by noting that ${\bf r_1}=a\hat{\bf r}-d_1\hat {z}$. So, the unit vector $\hat{\bf r}_1$ can be expressed as
\begin{align}\label{eq14}
\hat{\bf r}_1&=\frac{a\hat{\bf r}-d_1\hat{z}}{r_1}\nonumber\\
&=\frac{a\hat{\bf r}-d_1(\cos\theta\hat{\bf r}-\sin\theta\hat{\theta})}{r_1},
\end{align}
where we expressed $\hat{z}$ in terms of the $\hat{\bf r}$ and $\hat{\theta}$ of the coordinates centred at $O$. From the above equation, we can deduce the projection of $ \hat{\bf r}_1$ in the radial direction:
\begin{equation}\label{eq15}
    \hat{\bf r}_1\cdot\hat{\bf r}=\frac{a-d_1\cos\theta}{r_1}
\end{equation}
Using the cosine rule, we can express:
\begin{equation}\label{eq16}
\cos\theta_1=-\frac{d_1^2+{\bf r_1}^2-a^2}{2 d_1r_1}
\end{equation}
Further, expressing ${\bf r_1}^2$ in terms of $\theta$, it is possible to simplify $\cos\theta_1$:
\begin{align}\label{eq17}
\cos\theta_1=-\frac{d_1^2+(a^2+d_1^2-2ad_1\cos\theta)-a^2}{2dr_1}=-\frac{2d_1^2-2a\cos\theta}{2d_1r_1}=-\frac{d_1-a\cos\theta}{r_1}
\end{align}
Similarly, using the sine rule on the same triangle, one can deduce
\begin{align}\label{eq18}
    \sin\theta_1=\frac{a}{r_1}\sin\theta
\end{align}
The expression of $\hat{\theta}_1$ is tricky, as it is not immediately obvious how to proceed. Here we use the observation that ${\hat{\bf r}}_1\times{\hat{\bf\theta}}_1={\hat{\phi}}_1=\hat{\phi}\equiv\hat{\bf r}\times\hat{\bf\theta}$. Due to this, $\hat{\theta}_1$ can be expressed as
\begin{align}\label{eq19}
\hat{\theta}_1&=\hat{\phi}_1\times\hat{\bf r} _1=\hat{\phi}\times\hat{\bf r}_1
    =\hat{\phi}\times\frac{a\hat{\bf r}-d_1\hat{z}}{r_1}
\end{align}
Again, using the expression for $\hat{z}$, we get
\begin{align}\label{eq20}
\hat{\theta}_1
    &=\frac{1}{r_1}\left[(a-d_1\cos\theta)\hat{\theta}+d_1\sin\theta\hat{\bf r}\right]
\end{align}
On account of these results, the radial component of the magnetic field due to the source dipole at a point $(a,\theta)$ is given by
\begin{equation}\label{eq21}
{\bf B}^{source}(a,\theta)\cdot\hat{\bf r}=-\frac{\mu_0}{4\pi} m_1\left[\frac{2(d_1-a\cos\theta)(d_1\cos\theta-a)-ad_1\sin^2\theta}{(a^2+d_1^2-2ad_1\cos\theta)^{5/2}}\right]
\end{equation}
In the above analysis, we did not make use of any special property of the location of the source dipole being located outside the sphere. Therefore, one can find the radial component of the magnetic field at the surface point $(a,\theta)$ due to the image dipole:
\begin{equation}\label{eq22}
{\bf B}^{image}(a,\theta)\cdot\hat{\bf r}=-\frac{\mu_0}{4\pi} m_2\left[\frac{2(d_2-a\cos\theta)(d_2\cos\theta-a)-ad_2\sin^2\theta}{(a^2+d_2^2-2ad_2\cos\theta)^{5/2}}\right]
\end{equation}
We notice that by setting $d_2=a^2/d_1$, the denominator of Eq.\eqref{eq22} reduces to:
\begin{align}\label{eq23}
\left(a^2+\frac{a^4}{d_1^2}-2a\frac{a^2}{d_1} \cos\theta\right)^{5/2}=\frac{a^5}{d_1^5} \left(a^2+d_1^2-2ad_1\cos\theta\right)^{5/2}
\end{align}
On the other hand, using Eq.\eqref{eq11} and $d_2=a^2/d_1$, the numerator of 
Eq.\eqref{eq22} becomes:
\begin{align}\label{eq24}
-\frac{\mu_0}{4\pi}\left(-\frac{a^3}{d_1^3}m_1\right)\left[2(\frac{a^2}{d_1} -a\cos\theta)(\frac{a^2}{d_1} -a\cos\theta)-a\frac{a^2}{d_1}\sin^2\theta\right]\nonumber\\
=-\frac{\mu_0}{4\pi}\left(-\frac{a^5}{d_1^5}m_1\right) \left[2(a-d_1\cos\theta)(a-d_1\cos\theta)-ad_1\sin^2\theta\right]
\end{align}
Dividing Eq.\eqref{eq24} by Eq.\eqref{eq23}, we see that the factor ($a^5/d_1^5$) cancels, and we can find the fraction in Eq.\eqref{eq22} in terms of $m_1$, $a$, $d_1$ and $\theta$. Adding Eq.\eqref{eq21} and Eq.\eqref{eq22}, we can show that
\begin{align}\label{eq25}
\frac{\mu_0}{4\pi}m_1&\left[\frac{2(d_1-a\cos\theta)(d_1\cos\theta-a)-ad_1\sin^2\theta}{(a^2+d_1^2-2ad_1\cos\theta)^{5/2}}\right]\nonumber\\
&+\frac{\mu_0}{4\pi}m_2\left[\frac{2(d_2-a\cos\theta)(d_2\cos\theta-a)-ad_2\sin^2\theta}{(a^2+d_2^2-2ad_2\cos\theta)^{5/2}}\right]=0
\end{align}

Similarly, we can verify that the total vector potential on an arbitrary point $(a,\theta)$ on the sphere satisfies:
\begin{align}\label{eq26}
\frac{{\bf m_1}\times\hat{\bf r}_1}{r_1^2}+ \frac{{\bf m_2}\times\hat{\bf r}_2}{r_2^2}&= \frac{m_1\sin\theta_1}{r_1^2}(\hat{z}\times\hat{\bf r}_1)+\frac{m_2\sin \theta_2}{r_2^2}(\hat{z}\times\hat{\bf r}_2) \nonumber\\
&=\frac{m_1 a\sin\theta}{(a^2+d_1^2-2ad_1\cos \theta)^{3/2}}\hat{\phi}+\frac{m_2 a\sin\theta}{(a^2+d_2^2- 2ad_2\cos\theta)^{3/2}}\hat{\phi}={\bf 0}
\end{align}
when $d_2=\frac{a^2}{d_1}$ and $m_2=-\frac{a^3}{d_1^3}m_1$. Note that this shows that Eq.\eqref{eq2} holds true in this case. Eq.\eqref{eq25} and Eq.\eqref{eq26} validate all the boundary conditions. We note that the image dipole configuration is not liable to make sure that the total field (or vector potential) must be zero inside the sphere.
\section{Transverse dipole}
\begin{figure}[h]
\centering
\includegraphics[width=7.5cm, height=8.0cm]{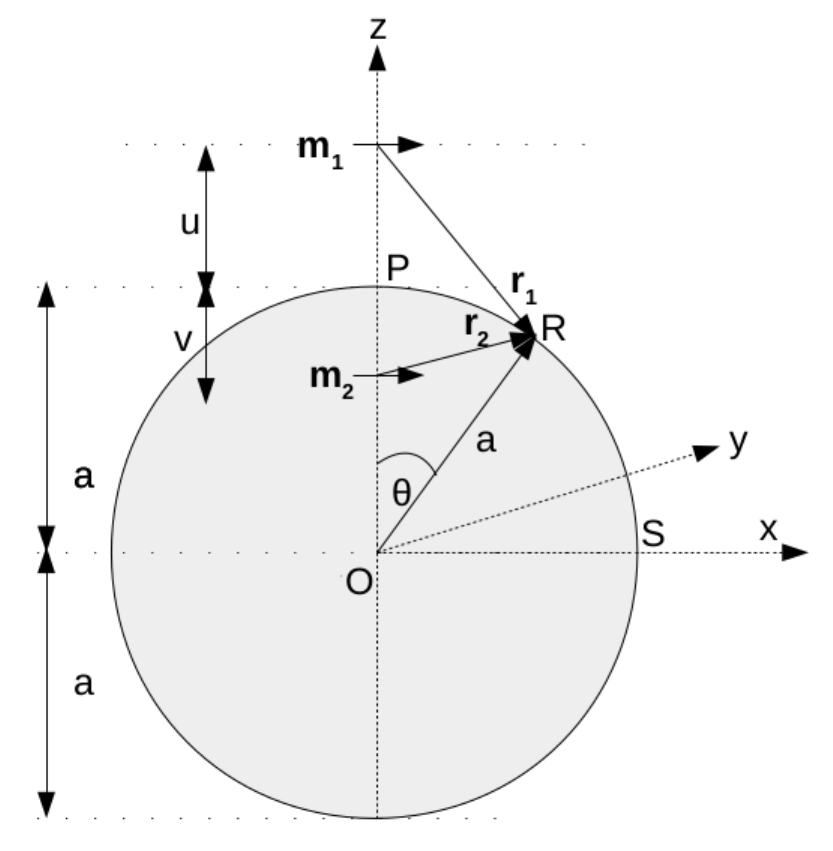}
\caption{Magnetic dipole placed with transverse orientation before a superconducting sphere}
\label{fig2}
\end{figure}

Now, we consider the case in which a source dipole is oriented in the transverse direction relative to the superconductor sphere. In figure~\ref{fig2}, we show that the dipole ${\bf m_{1}}$ is located on the $z$ axis at a distance $d_1$ from $O$ the centre of the superconducting sphere where $d_1>a$, 
the radius of the sphere. It points to the $x$ direction, i.e. transverse to the radial direction. The boundary condition is still given by Eq.\eqref{eq1} and Eq.\eqref{eq2}. Let us try to find a possible image configuration (i.e. the location and direction of the image dipole) using Eq.\eqref{eq1}. From the symmetry of the problem, it is clear that the image dipole must be conceived on the $z$ axis itself. If ${\bf m_{1}}$ is oriented along the $x$ axis, the image dipole (say ${\bf m_{2}}$) cannot be oriented along $y$ or $z$ axis for symmetry reasons. If ${\bf m_{2}}$ were indeed directed in the $z$ direction, then the radial component of the field due to ${\bf m_{1} }$, cannot cancel that due to ${\bf m_{2}}$ at all points on the sphere. Then, points would be found on the surface where the radial field component would remain nonzero. So, the image 
dipole may be aligned with the $x$ axis, as shown in Fig~\ref{fig2}. Further, we can see that there cannot be a single image point dipole in this case. This is because the radial components of the magnetic fields due to the source and image dipoles would not sum up to zero at the points under the location of the proposed image dipole. For 
example, consider the point $S$ where the $x$ axis punctures the sphere. In these regions, the field lines would be headed in similar directions. However, this does not happen as much at the points on the surface that are between the location of the image dipole and $P$, say at $R$, since in these points, the field lines are oppositely directed. Therefore, we find that another distribution of the magnetic dipole is required at the bottom of the originally proposed single dipole that would be directed in the opposite direction to compensate for the non-vanishing radial component of the field at the points located on the surface.
\subsection{Use of vector potential boundary condition}
To find the image configuration, we adopt another approach. We can always construct a magnetic dipole as a tiny current loop. Let us now treat a current loop as a successive set of straight current elements. We resort to the boundary condition of ${\bf A}$. Under the Coulomb gauge condition, the vector potential ${\bf A}$ satisfies $\nabla\cdot{\bf A}=0$. In this gauge, the vector potential satisfies a vector Poisson's equation and is given by:
\begin{equation}\label{Eq9}
    {\bf A}=\frac{\mu_0}{4\pi}\int\frac{{\bf I}dl'}{|{\bf r}-{\bf r'}|}
\end{equation}

The expression for the vector potential is akin to the expression for the electrostatic scalar potential except for the difference in the numerator where one would have an integral over a charge element instead of a current element. Otherwise, just like the scalar potential, the vector potential must be continuous across any surface. The key point to note here is that one may perhaps take the integral over the current element as a vector equivalent of the integral over a charge element. Inside the superconducting body the vector potential ${\bf}$ satisfies $\nabla\times{\bf A}={\bf0}$. As both the divergence and the curl of the vector field are zero in the simply-connected region inside the sphere, we can choose its value to be a constant (see the discussion after Eq.(7.4) and the beginning of section 7.2 in chapter 7 of~\cite{matsushita2021electricity}). When we combine this with the continuity of the vector potential across the surface, we can say that ${\bf A}$ is a constant on the surface. This resembles the condition of $\Phi=0$ on the surface of a conducting sphere for an electrostatic image problem. Beyond the surface, it assumes a functional form (note the kink in the vector potential in figures 7.4b, 7.7b, 7.8b of~\cite{matsushita2021electricity}). Specification of the boundary condition on ${\bf A}$ makes the solution to Laplace's equation unique. 
\begin{figure}[H]
    \centering
    \includegraphics[scale=0.5]{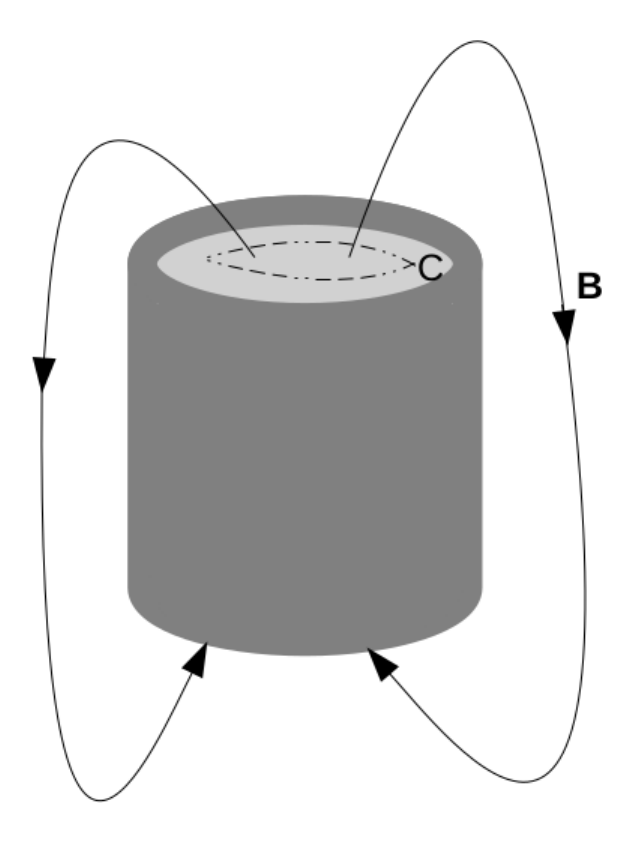}
    \caption{Trapping of magnetic field}
    \label{fig:enter-label}
\end{figure}

It is possible to see that in the case of a non-simply-connected region, e.g. inside a hollow superconducting cylinder, the magnetic vector potential is not constant. In this case, magnetic flux is trapped inside the hole and gets quantized in the units of $h/2e$~\cite{tinkham2004introduction}. Applying Ampere's law around a loop $\mathcal{C}$ lying completely within the hole, we find that:
\begin{align}
    \oint_{\mathcal{C}}{\bf A}\cdot d{\bf l}=\int{\bf B} \cdot d{\bf S}=n\frac{h}{2e},
\end{align}
where $h$ is Planck's constant, $e$ is elementary charge of an electron and $n$ is an integer. As a result, the vector potential depends on the radius of the loop. However, for a simply-connected superconductor, there is no trapping of magnetic flux. So perhaps we can take the value of the vector potential to be zero. In fact, we have already seen that for the radial orientation of the dipole, the vector potential adds up to zero (Eq.\eqref{eq26}) at the surface. In fact, if we choose the constant vector ${\bf A}={\bf 0}$ using the gauge freedom in ${\bf A}$, then it is possible to validate the boundary conditions in ${\bf B}$, as shown below. 

First, let us consider a tiny flat patch on the surface of the sphere. We construct a Gaussian pillbox of very small height on the patch. Then, the circumference of the patch acts as a boundary of the surface of the pillbox outside the sphere. Any non-zero normal component of the magnetic field would manifest as non-zero magnetic flux. If we apply Ampere's law to this surface and the loop, then the total magnetic flux coming out of the surface can be shown to be zero.
\begin{align}
    \Phi_B^\perp=\int{\bf B}\cdot d{\bf S}=\oint {\bf A}\cdot d{\bf l}=\oint {\bf 0}\cdot d{\bf l}=0
\end{align}
Because ${\bf A}={\bf 0}$ on the surface of the sphere. In the above equation, $d{\bf l}$ represents the differential length element around the circumference and $d{\bf S}$ represents the differential area element on the upper surface of the pillbox. This validates the boundary condition $B^ \perp=0$ (Eq.\eqref{eq1}). Similarly, we can construct another Amperian loop outside the superconductor whose one side grazes along the surface of the superconductor. In this case, the Gaussian pillbox is lying on the surface. Again, using the Ampere's law, we find that:
\begin{align}
    \Phi_B^{||}=\int{\bf B}\cdot d{\bf S}=\oint {\bf A}\cdot d{\bf l}=\int_{surface} {\bf 0}\cdot d{\bf l}+\int_{outside} {\bf A}\cdot d{\bf l}\neq0
\end{align}
Where the first line integral is evaluated on the $surface$ of the superconductor where we set ${\bf A}={\bf 0}$. But the other integral gives a non-zero contribution, since ${\bf A}$ is non-trivial outside the superconducting surface. This shows that the magnetic field must have components parallel to the surface.

Since the vector potential ${\bf A}$ is indeterminate up to the gradient of a gauge function $\chi$: ${\bf A} \rightarrow{\bf A}'={\bf A}+\nabla\chi$, therefore, it is safe to state that the ${\bf A}={\bf C}$ in a simply-connected superconductor. This justifies Eq.\eqref{eq2} referred to in~\cite{matsushita2021electricity}. In the following, we show the use of the boundary condition that the vector potential is a constant on the surface in the context of the problem where the source dipole is oriented transverse to the sphere. Mathematically, it implies that the differential of ${\bf A}$ (i.e. a very small change in ${\bf A}$) can be made arbitrarily close to zero at the surface: $d{\bf A}\rightarrow{\bf0}$. Since a given current element can always be written as a linear superposition of the radial and transverse elements, we can consider the two cases separately. 
\subsection{Image current elements}
In the following, we find the image current elements in transverse and radial directions.
\subsubsection{Image for transverse current element}
\begin{figure}[h]
    \centering
    \begin{subfigure}{0.5\textwidth}
    \centering
        \includegraphics[width=1.0\linewidth]{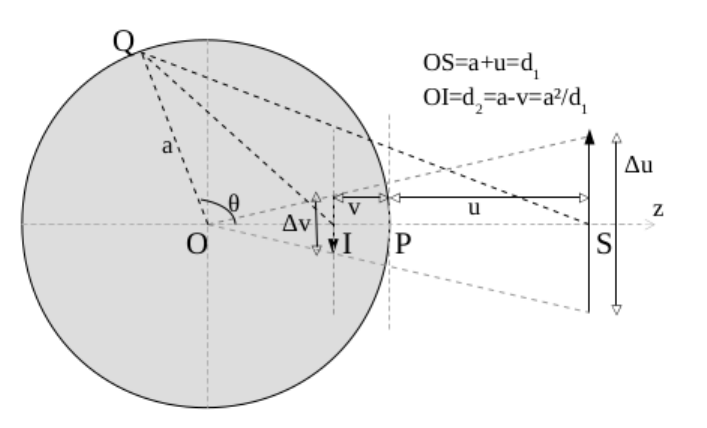}
        \caption{Transverse current element}
        \label{fig4a}
    \end{subfigure}%
    \begin{subfigure}{0.475\textwidth}
    \centering
        \includegraphics[width=1.0\linewidth]{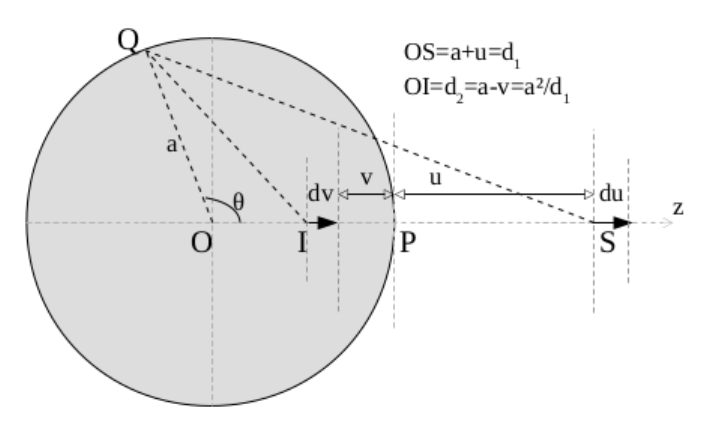}
        \caption{Radial current element}
        \label{fig4b}
    \end{subfigure}
\caption{Image current elements for (a) transverse and (b) radial orientation of source current element. The ${\bf S}$ource and the ${\bf I}$mage current elements are denoted by $S$ and $I$ in the figure ($I$ does not represent current). The length elements $\Delta u$ and $\Delta v$ are greatly exaggerated.}
\label{fig4}
\end{figure}

First, we consider the case of a source element carrying current $I_s^{(t)}$ is oriented in the transverse direction. It is passing through $S$ and has a length $\Delta u$, as shown in figure~\ref{fig4a}. The superscript $(t) $ stands for the transverse orientation of the current element. Let the image current element carrying image current $I_i^{(t)}$ have the length $\Delta v$ and is located at the point $I$ inverted with respect to the sphere. Let us say that the distance $OS=d_1$. Then, the image distance $OI=d_2=a^2/d_1$. Using the condition of constant vector potential on the surface, we can set:
\begin{align}\label{Eq11}
d{{\bf A}}_s^{(t)}&+d{{\bf A}}_i^{(t)}=0\nonumber\\
\implies \frac{{\bf I_s}^{(t)}{\Delta u}}{|\overrightarrow{OQ}-\overrightarrow{OS}| }&+\frac{{\bf I_i}^{(t)}{\Delta v}}{|\overrightarrow{OQ}-\overrightarrow{OI} |} =0\nonumber\\
\implies 
\frac{{\bf I_s}^{(t)}{\Delta u}}{\sqrt{a^2+d_1^2-2ad_1\cos\theta}}&+\frac {{\bf I_i}^{(t)}{\Delta v}}{\sqrt{a^2+ d_2^2-2ad_2\cos\theta}} =0\nonumber\\
\implies{\bf I_i}^{(t)}=-\frac{\sqrt{a^2+ (a^2/d_1)^2-2a(a^2/d_1)\cos\theta}}{\sqrt{a^2+ d_1^2-2ad_1\cos\theta}}\frac {\Delta u}{\Delta v}&{\bf I_s}^{(t)}=-\frac{a}{d_1}\frac{\sqrt{a^2+d_1^2-2ad_1 \cos\theta}}{\sqrt{a^2+ d_1^2-2ad_1\cos\theta}}\frac {\Delta u}{\Delta v}{\bf I_s}^{(t)}\nonumber\\
\implies{\bf I_i}^{(t)}=-\frac{a}{d_1}& \frac{d_1}{(a^2/d_1)}{\bf I_s}^{(t)}=-\frac{d_1}{a}{\bf I_s}^{(t)}
\end{align}

\subsubsection{Image for radial current element}\label{A-method}
Next, we consider the case of a source current element $du$ carrying current $I_s$, directed radially outwards and located at a distance $u=d_1-a$ from the pole $P$, as shown in figure~\ref{fig4b}. Comparing this configuration with the electrostatic case, we can conceive an image element $dv$ carrying current $I_i$, located at $v=\frac{a}{d_1}\left(d_1-a\right)$. To check the direction of that current we investigate the boundary condition of ${{\bf A}}^{(r)}$ (the superscript $(r)$ stands for the radial orientation of the current element) due to the source and image current elements:
\begin{align*}
    d{{\bf A}}_s^{(r)}&+d{{\bf A}}_i^{(r)}=0\nonumber\\
\implies \frac{{\bf I_s}^{(r)}\ du}{ |\overrightarrow{OQ}-\overrightarrow{OS}| }&+\frac{{\bf I_i}^{(r)}\ dv}{|\overrightarrow{OQ}-\overrightarrow{OI}| }=0\nonumber\\
\implies \frac{{\bf I_s}^{(r)}\ du}{\sqrt{a^2+d_1^2-2ad_1\cos\theta}}& +\frac{{\bf I_i}^{(r)}\ dv}{\sqrt{a^2+d_2^2-2ad_2\cos\theta}}=0\nonumber\\
\implies{\bf I_i}^{(r)}=-\frac{\sqrt{a^2+ (a^2/d_1)^2-2a(a^2/d_1)\cos\theta}}{\sqrt{a^2+ d_1^2-2ad_1\cos\theta}}&\frac {du}{dv}{\bf I_s}^{(r)}=-\frac{a}{d_1} \frac{du}{dv}{\bf I_s}^{(r)}
\end{align*}
Now, we know that as the source moves further, the image comes closer to the centre. In other words, $u$ and $v$ increase in opposite directions relative to the pole $P$. Hence $du/dv$ is a negative quantity. This was also pointed out in the context of longitudinal magnification in~\cite{bhattacharya2011analogy}. Its magnitude can be calculated from the mirror equation Eq.\eqref{eq10}: $|du/dv|=u^2/v^2=d_1^2/a^2$. As a result, the image radial current (along with direction) is given by:
\begin{equation}\label{Eq10}
    {\bf I_i}^{(r)}=-\frac{a}{d_1}\left(-\frac{u^2}{v^2}\right){\bf I_s}^{(r)}=\frac{d_1}{a}{\bf I_s}^{(r)}
\end{equation}

\subsubsection{Image for a radial current element from the closure of current loop}\label{Radial-image-2nd-method}
We can deduce the form of the image current element from the knowledge of the transverse current element's image and the closure of the current loops. 
\begin{figure}[H]
    \centering
    \includegraphics[scale=0.5]{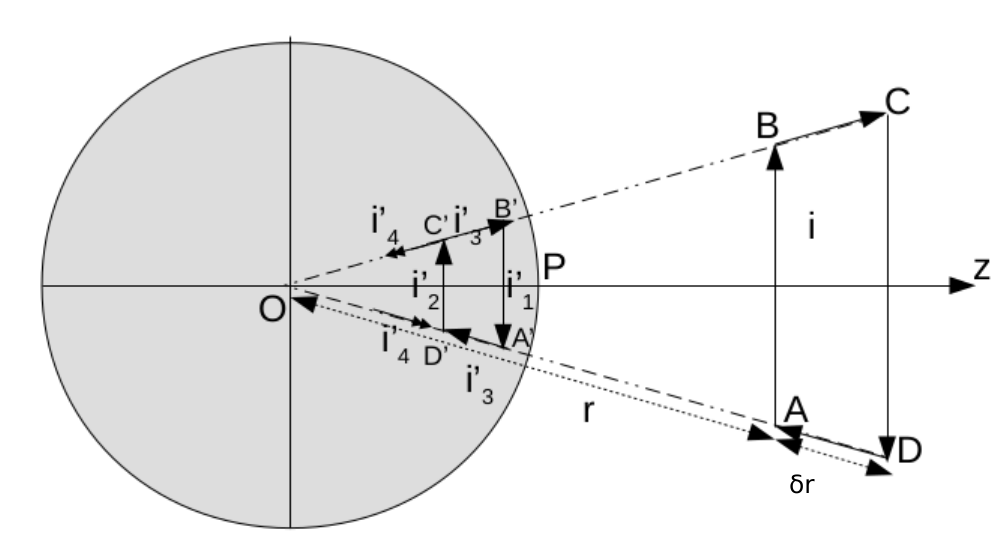}
\caption{The source and the image dipoles are represented by the current loops $ABCD$, and a combination of the loop $B'A'D'C'$ and the loop $C'OD'$. The image loops carry the current in opposite directions.}
    \label{fig5}
\end{figure}
In the above figure~\ref{fig5}, we consider a current loop $ABCD$ carrying current $i$ in the clockwise direction. We use the notation $r$ to denote the radial coordinate instead of $z$ and the notation $i$ to denote the current to keep it separate from the discussion in~\ref{A-method}. The images of the transverse elements of the arms $AB$ and $CD$ are $B'A'$ and $D'C'$ carrying current in the opposite direction. Using Eq.\eqref{Eq11} we can state that the magnitudes of the current $i'_1$ and $i'_2$ in the arms $B'A'$ and $D'C'$ are given by respectively:
\begin{subequations}
\begin{align}
    i'_1&=\frac{r}{a}i\\
    i'_2&=\frac{r+\delta r}{a}i
\end{align}
\end{subequations}
Now, the images of the radial elements must close the current loops, and by symmetry, themselves be radial. Segments $C'B'$ and $A'D'$ are the radial current elements that must carry current $i'_3$:
\begin{equation}\label{Eq12}
    i'_3=\frac{r}{a}i
\end{equation}
Because there are no other currents at the $A'$ and $B'$). Clearly, Eq.\eqref{Eq12} is identical to Eq.\eqref{Eq10}. The preceding analysis shows that irrespective of the radial or transverse orientation of the source current element, the image current element is scaled by the factor $\frac{r}{a}$ (or $\frac{d_1}{a}$ in the context of the figure~\ref{fig4b}). It is in the same direction as the source current for the radial case, and opposite of the source current in the transverse case.

The excess current $i'_4=\frac{\delta r}{a}i$ at the arm $D'C'$ must have come from the path $OD'$ at the junction $D'$ and must have gone towards $O$ at the junction $C'$. Thus, the only way to explain the excess current is to pass it through the centre along the path $C'OD'$. It appears due to our attempt to use a single image loop to satisfy the boundary conditions and the current closure condition. 

\subsection{Magnetic moment of the image dipole $A'D'C'B'$}
The first image dipole, as seen from the figure~\ref{fig5}, has the same direction as that of the source dipole. In the limit where the size of the current loop for the source dipole is vanishingly small, we can treat (i) the loop to represent a point dipole, (ii) the coordinate $r$ is almost along the $z$ direction, and (iii) the arms $AD$, $AB$ are nearly perpendicular to each other and the area of the loop will be given by the product $AD\cdot AB$. Corresponding arms of the image loop will be scaled as well. It is seen that the arm $A'D'$ will be scaled by a factor of $dv/du=a^2/r^2$. If the source dipole is located at $r=d_1$, we will have $A'D'=\frac{a^2}{d_1^2}AD$. Then, the arm $B'A'$ is scaled by a factor of $(a^2/d_1)/r=a^2/{d_1}^2$. As a result, the total area of the loop $B'A'D'C'$ is scaled down to $a^4/{d_1}^4(AD\cdot AB)$. To find the dipole moment, we must multiply this by the image current, whose magnitude is given by $\frac{d_1}{a}I$. Therefore, the dipole moment for this image dipole $|{\bf m_2}|=\frac{a^3}{{d_1}^3}|{\bf m_1}|$.




\subsection{Handling excess current}
The excess current $I''=\frac{\delta r}{a}I$ will flow through the triangular loop $C'OD'$. We can calculate the resulting dipole moment from the following figure~\ref{fig6a}.
\begin{figure}[H]
    \centering
    \begin{subfigure}{0.5\textwidth}
    \centering
        \includegraphics[width=1.0\linewidth]{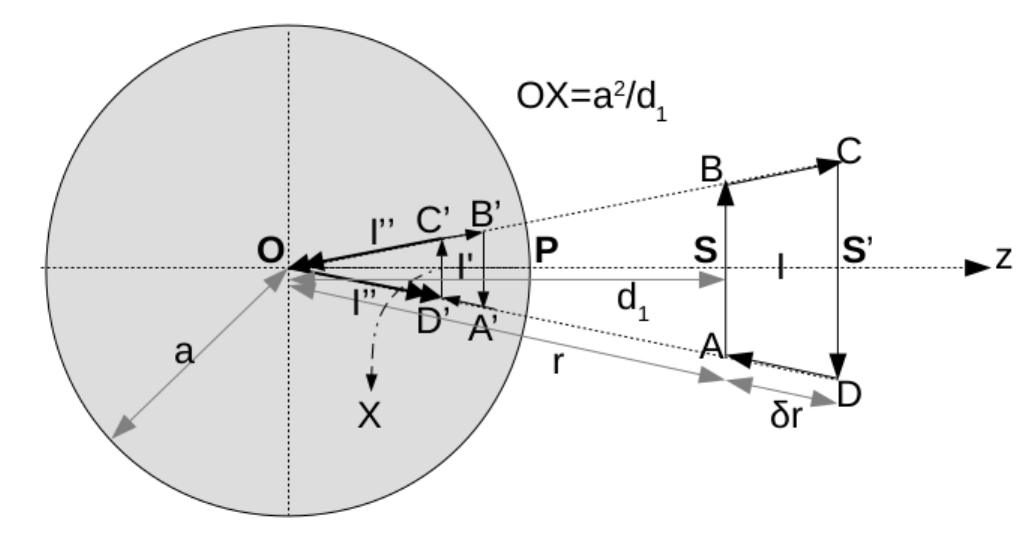}
        \caption{}
        \label{fig6a}
    \end{subfigure}%
    \begin{subfigure}{0.5\textwidth}
    \centering
        \includegraphics[width=1.0\linewidth]{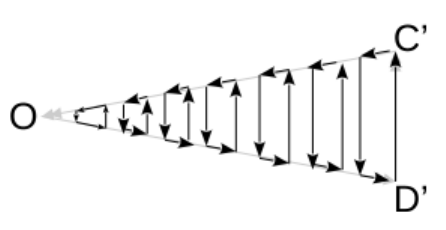}
        \caption{}
        \label{fig6b}
    \end{subfigure}
    \caption{(a) Excess current $I''=\frac{\delta r}{a}I$ passing through $O$ forms a dipole in the direction opposite to that created by the current in the image loop $A'D'C'B'$. The excess current was denoted by $i'_4$ in~\ref{Radial-image-2nd-method}; (b) Splitting up of current loop $C'OD'$ as an infinite series of tiny current loops representing linearly diminishing magnetic dipoles.}
    \label{fig6}
\end{figure}
The compensating dipole moment must be equal to the area of the triangle multiplied by the excess current:
\begin{align}\label{eq35}
    \mu&=\left(\frac{1}{2}C'D'\cdot OX\right)\frac{\delta r}{a}I\nonumber\\
    &\approx\left(\frac{1}{2}\frac{a^2}{{d_1}^2}AB\cdot \left(\frac{a^2}{{d_1}}\right)\right)\frac{\delta r}{a}I\nonumber\\
    &=\frac{a^3}{{d_1}^3}\left(\frac{1}{2}AB\cdot \delta r\cdot I\right)\nonumber\\
    &=\frac{a^3}{{d_1}^3}\frac{m_1}{2}
\end{align}
To check the consistency with the work by Lin~\cite{lin2006theoretical}, let us contemplate a linear magnetic dipole density of dipole moment per unit length $\lambda=\frac{d\mu'}{dz}=\frac{m_1}{ad_1}z$ along the $z$ axis between $O$ and $OX=a^2/d_1$. In that case, we can verify that the total dipole moment matches the above dipole moment due to compensating current.
\begin{align}\label{eq36}
    \mu'&=\int_0^{a^2/{d_1}}\frac{m_1}{a{d_1}}z\ dz\nonumber\\
    &=\frac{m_1}{a{d_1}}\left(\frac{z^2}{2}\right)^{a^2/{d_1}}_0\nonumber\\
    &=\frac{a^3}{{d_1}^3}\frac{m_1}{2}
\end{align}
Comparison of Eq.\eqref{eq36} with Eq.\eqref{eq35} shows the equivalence of this method with Lin's results.
\subsection{Validation of the result for radial dipole}
In this case, every point on the current loop representing the source dipole ${\bf m_1}$ is equidistant from the centre. Not only that, all elements are transverse to the superconducting sphere. As a result, the image current should be given by Eq.\eqref{Eq11}. Suppose the current loop for the source dipole is modelled as a tiny current loop located at a distance $d_1$ from the centre of the sphere on the $z$ axis. In that case, the length of the image dipole's current elements will be scaled down by a factor $\frac{a^2}{{d_1}^2}$. Assuming that the source current loop carries a current $I$, the magnetic dipole moment of the source dipole would be ${\bf m_1}=A{I}\hat{z}$, where $A$ represents the area of the loop, $A\rightarrow0$. On the other hand, the magnetic dipole moment for the image loop will be 
\begin{equation}
    {\bf m_2'}=-\frac{a^4}{{d_1}^4}A\cdot \left(\frac{d_1}{a}\right){I}\hat{z}=-\frac{a^3}{{d_1}^3}(A{I}\hat{z})=-\frac{a^3}{{d_1}^3}{\bf m_1}
\end{equation}
This validates the fact that the model based on current loops is effective in explaining both the radial and transverse orientation of the magnetic dipole placed in front of a superconducting sphere.
\section{Summary}
In this work, we solved the problem of a magnetic dipole placed in front of a superconducting sphere using the method of images in a way accessible to undergraduate physics courses. The case of the radial dipole was solved using the boundary condition that $B^\perp=0$. The case of the transverse orientation of the source dipole has been studied by Lin~\cite{lin2006theoretical} in the past. We approached the problem through an intuitive pathway of representing magnetic dipoles by current loops and finding images of the elements of the current loop. Invoking the similarity between the vector potential and the electrostatic potential under the Coulomb gauge, and the boundary condition on vector potential, we determined the image of the transverse current element. Further, we deduced the image of the radial current element using two different methods: (a) boundary condition of ${\bf A}$ and (b) loop closure method. Using these results, we determined the image configuration of a current loop that represents a source magnetic dipole. This result was shown to be consistent with Lin's work and also with the case of a radially oriented source dipole. So, the novelty of the present work is not in discovering the current element method, but in demonstrating that the method can be used to solve the problem where the dipole is oriented transversely relative to the sphere. This task is not trivially accomplished, since it requires conceiving a continuous distribution of image magnetic dipoles down to the centre of the sphere. But the treatment in section~\ref{Sec2} is also useful due to its elegance and the practice of vector algebra. Additionally, it presents an alternative way of solving the problem. Overall, this example of the method of images involving superconductors is interesting from a pedagogical perspective because the source is a vector quantity, unlike the grounded conducting sphere image problem. Moreover, it presents us with the unique feature of a continuous image dipole distribution. This does not happen when an electric dipole is placed near a conducting sphere~\cite{santos2004electrostatic}. For completeness, let us mention that if we place the magnetic dipole in front of a sphere made of normal conductor (like Aluminium or Copper) instead of a superconductor, then the boundary conditions Eq.\eqref{eq1} and Eq.\eqref{eq2} cannot be used. The relevant boundary conditions should be:
\begin{enumerate}
    \item electric field should vanish inside the conductor
    \item Normal component of the magnetic field and the tangential component of the electric field should be continuous across the surface.
\end{enumerate}
The method of images can still be used if charges change value or position slowly relative to the relaxation time of the conductor~\cite{staelin2011electromagnetics}. See the qualitative discussion related to Fig. 4.2.3 in section 4.2 of~\cite{staelin2011electromagnetics}. For superconductors, we can solve a problem as shown in this paper. We expect that this work will help the teaching-learning process and will gain the appreciation of the students and teachers.

\section{Importance in physics pedagogy}
The teachers who teach intermediate and advanced electricity and magnetism courses for undergraduate or master's level students will benefit from this work. One limiting factor in this work is that superconductivity is not a common undergraduate topic. However, with some basic introduction to electromagnetism of superconductors, as shown in Matsushita~\cite{matsushita2021electricity}, this work builds up an analogue to an electrostatics problem. It provides a version of the solution that is more undergraduate-ready than the cited literature, adding pedagogical appeal to the work. 

\section{Conflict of interests}
The authors declare no conflict of interests, financial or otherwise, with any known party.
\section{Data availability statement}
No new data was generated or analysed in this study.
\section{Acknowledgements}
We would like to thank the National Initiative on Undergraduate Science (NIUS) Program, run by the Homi Bhabha Centre for Science Education for supporting this research. No funding was received to carry out this work.
\section{References}
\bibliographystyle{unsrt}
\bibliography{MagDipole}
\end{document}